\documentclass[useAMS,usenatbib]{mn2e}
\usepackage{graphicx}

\title[The many lives of active galactic nuclei-II]{The many lives of active galactic nuclei-II: The formation and evolution of radio jets and their impact on galaxy evolution}

\author[M. Raouf et al.]{Mojtaba Raouf $^{1,2}$\thanks{E-mail:
		m.raouf@ipm.ir},
	Stanislav S. Shabala $^{3}$,
	Darren J. Croton $^{2}$,
	Habib G. Khosroshahi $^{1}$,
	\newauthor
	Maksym Bernyk $^{2}$.
	\\
	$^{1}$School of Astronomy, Institute for Research in Fundamental Sciences (IPM),
	Tehran, 19395-5531, Iran\\
	$^{2}$ Centre for Astrophysics \& Supercomputing, Swinburne University of Technology, PO Box 218, Hawthorn, Victoria 3122, Australia \\
	$^{3}$School of Physical Sciences, Private Bag 37, University of Tasmania, Hobart, TAS 7001, Australia
}

\begin{document}
	
	\date{}
	\pagerange{\pageref{firstpage}--\pageref{lastpage}} \pubyear{2012}
	\maketitle
	\label{firstpage}

	\begin{abstract}
		We describe new efforts to model radio active galactic nuclei (AGN) in a cosmological context using the SAGE semi-analytic galaxy model. Our new method tracks the physical properties of radio jets in massive galaxies, including the evolution of radio lobes and their impact on the surrounding gas. This model also self consistently follows the gas cooling-heating cycle that significantly shapes star formation and the life and death of many galaxy types. Adding jet physics to SAGE adds new physical properties to the model output, which in turn allows us to make more detailed predictions for the radio AGN population. After calibrating the model to a set of core observations we analyse predictions for jet power, radio cocoon size, radio luminosity, and stellar mass. We find that the model is able to match the stellar mass--radio luminosity relation at $z\sim0$, and the radio luminosity function out to $z\sim1$. This updated model will make possible the construction of customised AGN-focused mock survey catalogues to be used for large-scale observing programs.
	\end{abstract}


	\section{Introduction} 
	
	The paradigm of hierarchical galaxy formation provides a remarkably successful framework to explain many properties of the observed galaxy population, particularly at intermediate masses \citep{Press1974,White1978,White1991,Lacey1994}. With early models over-predicting galaxy counts at both the bright and faint end of the luminosity function, it was quickly realised how important feedback processes were for reproducing the observations. At the low mass end, heating due to the reionization of the Universe at early times, and supernovae ejecta from the most massive stars, became critical to explain the deficit of faint galaxies with respect to the number of low mass halos \citep{Efstathiou1992}. However in the most massive haloes such feedback modes proved insufficient to offset rapid cooling of the hot gas, leading to a subsequent overproduction of massive elliptical galaxies by many orders-of-magnitude.\footnote{The ratio of galaxy stellar mass to dark matter halo mass decreases with increasing mass. Hence, even at a constant specific star formation rate the efficiency with which supernova can eject gas out of the gravitational potential decreases with mass. In reality, massive ellipticals have lower specific star formation rates than spirals, and the impact of the supernova is even smaller.} Furthermore, optical observations showed these massive ellipticals to have formed only a small fraction of their stars over the last half of the Hubble time \citep{Bender1999}, in sharp contrast to the models which predicted that these objects should be rapidly forming stars at the present epoch.
	
	A closely related problem was presented by galaxy clusters. Observations of X-ray emission suggested that there should be runaway cooling of the hot ($\sim 10^7$K) gas \citep[the so-called ``cooling catastrophe'';][]{Cowie1977,Fabian1977}, however no such rapidly cooling gas at temperatures below about a third of the virial temperature was found \citep{Tamura2001,Peterson2003}. A new mechanism was required to offset gas cooling on scales of tens of kpc, and suppress excessive galaxy growth from star formation.
	
	\cite{Silk1998} pointed out that the presence of a supermassive black hole (SMBH) at the centre of every massive galaxy provides such a mechanism, through conversion of a fraction of accreted mass to thermal or kinetic energy \citep{Penrose1971,Blandford1977,Blandford1982}; this is the mechanism invoked to explain the presence of Active Galactic Nuclei (AGN). By coupling a fraction of the AGN energy output to the rapidly cooling hot halo gas, a new generation of galaxy formation models \citep{Granato2004,Monaco2006,Croton2006,Bower2006} succeeded in reproducing the quenching of late epoch star formation in the most massive galaxies. While successful in the sense of reproducing the observed galaxy properties, these models do not make detailed predictions for the mechanisms through which AGN energy couples to the gas. 

	Broadly speaking, AGN feedback can be either mechanical (through radio jets) or radiative \citep[see reviews by][]{Cattaneo2009,Somerville2015}, with radiative feedback affecting the ISM of the host galaxy, and radio jets doing feedback on the larger halo scales. For this reason, \cite{Croton2006} referred to the mode of AGN feedback responsible for suppressing star formation since $z\sim1$ as ``radio'' or ``maintenance'' mode feedback. Dramatic evidence for this mode of feedback in action was found in the Perseus \citep{Bohringer1991,Fabian2003} and Virgo \citep{Churazov2001,Forman2005} clusters, where lobes of radio emitting plasma were observed to displace the X-ray emitting gas away from the rapidly cooling central regions.
	
	Observationally, the radio loud AGN fraction is found to be a strong function of host galaxy properties: \cite{Sadler1989} found more than 40\% of massive ellipticals to host low-luminosity radio sources. \cite{Best2005} showed that this fraction scales strongly with stellar mass, and this scaling is consistent with a picture of ``maintenance mode'' feedback, in which massive galaxies with rapidly cooling hot haloes are more likely to host AGN, which can in turn provide the feedback needed to offset the otherwise imminent cooling catastrophe. In clusters, the state of the cooling gas appears to be intimately connected to the probability of finding an AGN in the cooling gas: \cite{Burns1990} reported that over 70\% of cool core clusters hosted radio sources, compared to only 23\% of non-cool core clusters. Other authors \citep[e.g.][]{Mittal2011} have found similar results.
	
	While many of these sources are compact \citep{Shabala2008,Sadler2014}, when the jets are powerful enough they can break through the galaxy disk and inflate pairs of lobes of radio emitting synchrotron plasma, with the largest lobes up to a Mpc in size \citep[][see  \cite{Banfield2016} for an example of a recent discovery]{Laing1983,Saripalli1986}. Morphologically, extended radio lobes are observed to be either core or edge-brightened; such sources are classified as Fanaroff-Riley type I and type II, respectively \citep{Fanaroff1974}. 
	
	Large radio sources can do feedback well outside the host galaxy disk via two channels. First, powerful radio sources can drive strong shocks through the surrounding gas \citep{Schoenmakers2000,Rawlings2004,Shabala2011}, heating and uplifting it to large radii. Second, lower power sources in galaxy clusters are often observed to give rise to buoyant bubbles of radio plasma, which displace the hot X-ray emitting gas as they rise through the cluster \citep{Bohringer1991,Churazov2001,Fabian2003,Forman2005}.
	
	Multiple lines of evidence suggest that radio AGN activity must be episodic. Theoretically, AGN intermittency is expected in a self-regulating feedback process \citep[e.g.][]{Bahcall1997,Kawata2005,Novak2011,Gaspari2015}. Observationally, the presence of multiple shocks and ripples around the X-ray cavities in the Perseus \citep{Fabian2003} and Virgo \citep{Forman2005} clusters have been interpreted as remnants of multiple AGN outbursts. Perhaps the most dramatic evidence for recurrent radio AGN activity is provided by sources with multiple pairs of radio lobes \citep[e.g.;][]{Schoenmakers2000}. In these ``double-double'' radio sources, the inner pair of lobes is interpreted to correspond to the most recent AGN outburst, while the outer pair of lobes is due to a previous episode of AGN activity. In a number of sources \citep[e.g. Centaurus A;][ a similar suggestion has also been been made for Cygnus A, Chon et al. 2012]{Israel1998,Feain2009}, the jet directions are often misaligned between outbursts, providing a mechanism for isotropizing the coupling of jet energy to cluster gas.
	
	The interaction between radio jets / lobes and the hot atmospheres into which they expand sets both the properties of the observed AGN, such as their sizes, morphologies, luminosities and radio spectra; and, through feedback, those of their host galaxies. The effect of environment on both AGN properties and the amount of feedback they do has been extensively studied through radio source dynamical models \citep{Scheuer1974,Begelman1989,Kaiser1997,Alexander2002,Shabala2008,Turner2015}, and numerical simulations \citep{Norman1982,Reynolds2002,Basson2003,Krause2005,Mendygral2012,Hardcastle2013,Hardcastle2014}. Recently, \cite{Godfrey2016} pointed out that the often quoted relationship between jet kinetic power and radio luminosity \citep[derived from observations of X-ray deficient cavities, e.g.;][]{Birzan2004,Birzan2008,Cavagnolo2010} is in fact driven by strong selection effects, and the true relationship is more complicated due to environmental effects \citep[e.g.][]{Barthel1996,Kaiser1997,Hardcastle2013,Hardcastle2014}. Observationally, \citet{Khosroshahi2017} found that quantities such as the radio luminosity of the brightest group galaxies strongly depend on their environment, such that the brightest group galaxies in dynamically young (evolving) groups are an order-of-magnitude more luminous in the radio than those with a similar stellar mass but residing in dynamically old (evolved) groups \citep[for a definition of old and young groups see][]{Raouf2014}. This finding is consistent with results of hydrodynamical simulations \citep{Raouf2016}, which suggest that the IGM in dynamically evolved groups is hotter for a given halo mass than that in evolving groups.
	
	Despite the importance of feedback from radio jets, to date no galaxy formation model has attempted to predict simultaneously the properties of both galaxies and AGN in detail. \cite{Fanidakis2012} produced an AGN radio luminosity function, however they employed both an arbitrary scaling between jet power and radio luminosity, and an arbitrary normalisation of the radio luminosity function; they therefore did not properly account for the intermittency of the feedback process. Shabala \& Alexander (2009) used a dynamical radio source model to quantify the feedback from jets on the hot gas, however they made no predictions for the resulting AGN properties. 

	A major difficulty in connecting jet models and simulations with observations lies in quantifying the environments into which the jet expands. Density and temperature profiles derived from X-ray observations are the gold standard, however these are typically biased towards low-redshift dense environments. \cite{Turner2015} introduced a model which connects radio AGN dynamical models with semi-analytic models of galaxy formation. In this approach, the semi-analytic model was used to estimate the total hot gas mass contained with the virial radius of a given dark matter halo; this gas was then assumed to follow a density profile consistent with observations of local clusters \citep[e.g.;][]{Vikhlinin2006}. \cite{Turner2015} used this approach to derive the physical properties (jet powers and ages) of radio AGN in a volume-limited low-redshift sample and found it reproduces well a number of key AGN observables. Here, we adopt their techniques to develop a new prescription for intermittent AGN feedback in the SAGE (Semi-Analytic Galaxy Evolution) galaxy formation model \citep{Croton2016}, updating the more simplistic ``radio mode'' model introduced in \citet{Croton2006}. Our code is publicly available as a fork of the original SAGE repository.\footnote{https://github.com/mojtabaraouf/sage}
	
	This paper is organised as follows: In Sections 2 and 3 we describe our N-body and semi-analytic framework, respectively. In Section 4 we describe the major features of our model for AGN feedback. Model constraints and predictions are discussed in Section 5. We present the summary of our results in Section 6. 

	\section{The dark matter simulation: Millennium}
	\label{Sec:MS}
	
	In this work we use the Millennium Simulation \citep{Springel2005} N-body dark matter halo merger trees as input into the ``Semi-Analytic Galaxy Evolution'' (\textsc{SAGE}) galaxy formation model \citep{Croton2016}, which we update with new AGN physics (described below). This simulation is important as it provides the structural backbone onto which galaxies (and hence black holes) can be evolved.
	
	The Millennium Simulation was run using the popular \textsc{GADGET-2} code and adopted a cosmological model consistent with the first year Wilkinson Microwave Anisotropy Probe data \citep[WMAP-1,  with parameters $\Omega_m = 0.25$, $\Omega_{\Lambda} = 0.75$ and ${\rm H_0} = 100 {\rm h}$ km $s^{-1}$ $\rm Mpc^{-1}$ where $\rm h=0.73$]{Spergel2003}. The simulation box of $(500 \rm h^{-1} \rm Mpc)^{3}$ contained $2160^{3}$ particles and had a mass resolution of $8.6 \times 10^8 \rm h^{-1} M_{\odot}$ per particle. 64 snapshots of the particle evolution were written to disk, spaced approximately logarithmically in scale factor between ${z=127}$ and ${z=0}$. Dark matter halos were then found amongst the mass distribution in each output, characterised as associations of 20 or more bound particles. For this, a combination of the Friends-of-Friends \citep{Davis1985} and SUBFIND \citep{Springel2001} halo finding algorithms were applied. After all dark matter halos had been identified they were then linked into their respective merger trees using the \textsc{L-HALOTREE} code. From this the full growth history of each ${z=0}$ object in the box could be inferred. For more information on the Millennium Simulation see \cite{Springel2005}.
	
	Note that the exact simulation and cosmology used is somewhat secondary for the goals of this paper. \textsc{SAGE}, and our new AGN model coupled to it, can be run on any simulation, and the model parameters allow it to be calibrated to match key observations in physically sensible ways.  
	
	\section{The galaxy formation model: SAGE}
	\label{Sec:SAGE}
	
	We only give a brief introduction to the base \textsc{SAGE} galaxy formation model here and refer the interested reader to \cite{Croton2016} for a full description. Beyond this, the rest of the paper will focus on the model changes, primarily related to supermassive black hole growth and outflows, that lead to a more physically motivated coupling between AGN feedback and galaxy evolution. This includes the key observables those working in this area of research might want such a model to predict.
	
	\textsc{SAGE} is an updated version of the semi-analytic model first introduced in \cite{Croton2006}. It analytically follows the movement of baryons through different mass reservoirs, computed on top of the numerically determined evolving dark matter halo mass distribution, characterised by a set of N-body simulation halo merger trees. Baryonic reservoirs include the hot halo gas, cold disk gas, stars in the disk, black holes, and gas ejected from the halo due to feedback events. \textit{How} this mass moves between the reservoirs is determined by a series of coupled differential equations that describe each physical process believed important. These include hot gas cooling into the disk (hot$\rightarrow$cold), star formation (cold$\rightarrow$ stars), gas heating from supernova or AGN feedback (cold$\rightarrow$ hot), ejection (cold/hot$\rightarrow$ ejected) and later reincorporation (ejected$\rightarrow$ hot), the effects of reionization, and so on. Most processes have one or more efficiency parameters that allow us to individually control their relative importance, although in reality the different components of galaxy and AGN evolution are highly intertwined and not so easily broken up  \citep[see e.g.][]{Mutch2013}.
	
To ensure \textsc{SAGE} produces a galaxy population akin to that observed around us it is calibrated by hand to statistically match a set of key observables. Our primary observable is the local stellar mass function, with a set of secondary observables being the star formation rate density history, net cooling rate--temperature relation and the AGN radio luminosity function (part of the new model extension),  shown and discussed below in Section \ref{Sec:Constrains}. All model results assume a Universe where ${h = 0.73}$, and when relevant, a Chabrier initial mass function \citep{Chabrier2003}  to compare model to observed stellar masses.

	\begin{figure}
		\centering
		\includegraphics[width=0.45\textwidth]{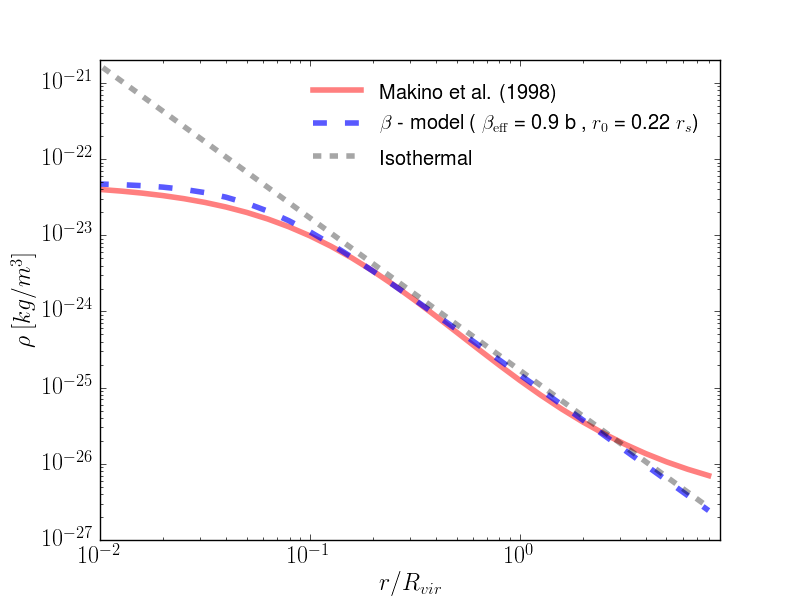}
		\caption{The hot gas density profile as a function of $r/R_{\rm vir}$ from Makino et al. (1998) (red solid line) and the original SAGE isothermal profile (gray short -dashed line). Densities are normalized to have the same total hot gas mass ($m_{\rm hot} = 10^{13} \rm h^{-1} M_{\odot}$). The best fit $\beta$-profile, with $\beta_{\rm eff} = 0.9 b$ and $r_0 = 0.22 r_s$, is also plotted with the  blue long-dashed line for comparison.}
		\label{fig:Densityp}
	\end{figure}
	
	\begin{table*}
		\centering
		\scriptsize
		\caption{The new AGN model parameters, their best values, plausible ranges, equation number(s), section(s), and a brief description, as used throughout this paper.}
		\label{tab:value}
		\begin{tabular}{llllll}
			\hline \\	
			Parameter	         &   Best value     & Plausible range  & Equation(s) & Section(s) & Description\\ \hline \hline	
			&&&&\textbf{Environment}\\ \hline	
			$\beta_{\rm eff}$              &   0.9 $b$   &    $\leq$0.66 &  $\ref{eq:betafit}$ & $\ref{sec:Densityprofile}$& Effective $\beta$ in \cite{Makino1998} \\	
			$\beta$              &   3($\beta_{\rm eff}$)   &    $\leq$ 2  &  $\ref{eqn:Rcocoon}$, $\ref{eqn:a_D}$, $\ref{eqn:lamdashock}$,$\ref{eqn:Tshock}$, $\ref{eqn:u_B}$,$\ref{eqn:f_p}$ & $\ref{Sec:RadioExpan}$,$\ref{Sec:Rshock}$, $\ref{Sec:GasNewHot}$,$\ref{Sec:JetProper}$&Slope in the $\beta$ density profile\\
			$r_0$                &   0.22 $r_s$ & fixed & $ \ref{eq:betafit}$,$\ref{eqn:Rcocoon}$,$\ref{eqn:Mshock}$,$\ref{eqn:u_B}$, $\ref{eqn:Lradio}$   & $\ref{sec:Densityprofile}$,$\ref{Sec:RadioExpan}$,$\ref{Sec:GasNewHot}$,$\ref{Sec:JetProper}$& Core radius in the  \cite{Makino1998} density profile\\
			\hline
			&&&&\textbf{Lobe dynamics}\\ \hline
			$\Gamma_{\rm x}$	 &   5/3    &     fixed & $\ref{eqn:a_D}$  & $\ref{Sec:RadioExpan}$&ICM adiabatic index\\
			$\Gamma_{\rm l}$	 &   4/3    &     fixed & $\ref{eqn:a_D}$  & $\ref{Sec:RadioExpan}$&Radio lobe adiabatic index\\
			$\eta$               &   0.35  &  0.001-1 &  $\ref{eqn:Qjet_eta}$   & $\ref{Sec:RadioJet}$& Jet generation efficiency\\
			$\kappa_R$                &   1.0  &  0.0-1.0 & $\ref{eq:BHaccr}$ & $\ref{Sec:RadioJet}$& Radio mode feedback efficiency \citep{Croton2006} \\
			
			$a_D$     &   1.0  & 0.8--1.8 & $\ref{eqn:Rcocoon}$,$\ref{eqn:a_D}$,$\ref{eqn:f_p}$   &$\ref{Sec:RadioExpan}$,$\ref{Sec:JetProper}$& Dimensionless constant related to radio source size \citep{Alexander2002}\\
			$R_{\rm T}$	         &   6    &     1.3--6  & $\ref{eqn:a_D}$,$\ref{eqn:f_p}$,$\ref{eqn:Lradio}$   & $\ref{Sec:RadioExpan}$,$\ref{Sec:JetProper}$& Cocoon axial ratio\\ 
						
			\hline	
			&&&&\textbf{Radio emission}\\ \hline
			$\nu$            	 &   1.4 ~GHz&    --  & $\ref{eqn:Lradio}$  &$\ref{Sec:JetProper}$& Rest-frame observing frequency\\
			$p$ 	             &   2.6     &    2.1-2.7 &  $\ref{eqn:Lradio_u_B}$,$\ref{eqn:A_1}$,$\ref{eqn:C_2P}$, $\ref{eqn:Lradio}$ & $\ref{Sec:JetProper}$& Power-law electron injection index\\
			
			$\gamma_1$ 	         &   1      &     1--$10^3$ &  $\ref{eqn:n_gama}$  & $\ref{Sec:JetProper}$& Minimum Lorentz factor\\
			$\gamma_2$	         &   $10^6$ &     $10^5$--$10^6$ &  $\ref{eqn:n_gama}$ & $\ref{Sec:JetProper}$& Maximum Lorentz factor\\
			
			$C_2(p)$           &   $2.04 \times 10^{-3}$ & fixed & $\ref{eqn:C_2P}$ & $\ref{Sec:JetProper}$& Constant derived from $p$ \\
			$A1$     &   $2.9 \times 10^{-9}$  & fixed & $\ref{eqn:A_1}$   & $\ref{Sec:JetProper}$& Constant derived from $p$\\
			\hline \\
		\end{tabular}
		
	\end{table*}

	\section{A new model of AGN feedback in SAGE}
	
	The impact of an AGN jet on the surrounding gas outside of a galaxy was captured by the model of \cite{Shabala2009}, who calculated the dynamics of jet-inflated radio lobes that propagate supersonically and shock heat the surrounding gas. Their work is applicable both for ``hot mode'' and ``cold mode'' black hole accretion that produces AGN jets with different efficiencies, and hence lends itself nicely for incorporation into \textsc{SAGE}. In the hot mode, also called the radio mode \citep{Croton2006}, accretion occurs at low Eddington rates and the accretion disk is geometrically thick and optically thin. This results in longer cooling times and more powerful jets. In contrast, the cold mode is associated with high Eddington rates and is described via the standard Shakura-Sunyaev geometrically thin, optically thick disk. The cold mode is radiatively efficient and produces a quasi-black-body spectrum, including optical / UV continuum and narrow-line AGN emission.
	
	Both radio lobe dynamics and associated feedback are sensitive to the environment into which the lobes expand \citep{Kaiser1997,Willott1999,Turner2015}. For self-consistently between the new AGN and existing galaxy model, we therefore first need to update the \textsc{SAGE} hot halo gas density profile into which the jets propagate. This in turn has an effect on the hot gas cooling rates into the galaxy. We then analytically describe the AGN jet evolution itself and how this replaces the existing more simplistic \textsc{SAGE} radio mode feedback. 
	
	We note that all figures and results presented in this paper arise from this new model using the parameters given in Table \ref{tab:value}. These parameters are defined in the various subsections below when they are introduced. Control over the model behaviours  is achieved through  parameterizing its different components, essentially to  set  their   relative efficiency, as (currently)  no model of galaxies  can be described from first principles alone. Such ``tuning" also accounts for the many finer details that are ill-understood and can not be easily modeled explicitly. Parameter values must be consistent with the physical process they are meant to describe and are set so that the overall model agrees with a key set of galaxy and AGN observations. Beyond a certain point, adding more parameters typically makes the model harder to calibrate.

	\subsection{The hot gas density profile and cooling} 
	\label{sec:Densityprofile}
	
	In \textsc{SAGE} every dark matter halo is assumed to carry its cosmic share of baryons, taken to be $f_b = 0.17$, consistent with the WMAP1 results of \cite{Spergel2003}. These baryons begin their life as diffuse hot gas with primordial composition around the galaxy. However, as mentioned above, with time the gas transforms under the action of the many and varied physical processes of galaxy evolution to populate gas reservoirs of several different phases, as well as stars, black holes, and the heavy elements \citep{Baugh2007}.
	
	Considering first this hot gas in virial equilibrium, its temperature can be described by \citep{Sutherland1993} 
	\begin{equation}
		T_{\rm vir} = \frac{1}{2} \frac{\mu m_p}{k_B} V_{\rm circ}^2 ~,
		\label{eqn:Tvir}
	\end{equation}
	where $m_p$ is the mass of proton, $\mu$ is the mean molecular weight of gas, $k_B$ is the Boltzmann constant, and $V_{\rm circ}$ is the circular velocity at the virial radius of halo. In what follows we approximate $V_{\rm circ}$ by $V_{\rm vir}$ of the halo. Of course, after an AGN jet ploughs through this gas it is unlikely to be in equilibrium, at least for a time. The new hot temperature of such gas is discussed in the following sections by taking into account the size and properties of an expanding shocked gas bubble due to an AGN outflow. 
	
	The hot gas density profile is similarly important as it directly determines the cooling (i.e. feeding) rate of gas into the galactic disk, gas which ultimately leads to the formation of new stars. The original \textsc{SAGE} model, like many other semi-analytic models before it, assumed that this hot gas can be represented as a simple isothermal sphere with temperature given above, and having a density profile \citep{White1991}
	\begin{equation}
		\rho(r) = \frac{m_{\rm hot}}{4 \pi R_{\rm vir} r^2} ~.
		\label{eq:isothermal}
	\end{equation}
	Here $m_{\rm hot}$ is the total hot gas mass in the halo within the virial radius, $R_{\rm vir}$. 
	
	However this approximation is clearly an oversimplification, especially when compared to X-ray observations of nearby cluster systems which show that the density profile is better described by an isothermal $\beta$-model \citep{Cavaliere1978,Fukazawa2004,Pointecouteau2004,Vikhlinin2006},

	\begin{equation}
		\rho(r) = {\rho_{0} \over [1+(r/r_c)^2]^{3\beta/2}} ~,
		\label{eq:betaprofile}
	\end{equation}
	where $r_c$ defines a characteristic core radius, and $\beta$ describes the changing inner and outer slopes of the profile.
	
	For the purpose of the present study we adopt the density profile described by \cite{Makino1998}. The Makino profile closely matches the observationally fit $\beta$-model given by Equation~\ref{eq:betaprofile}, albeit based on a better theoretical foundation. It characterises the mass distribution that arises from isothermal gas in hydrostatic equilibrium analytically embedded in a universal ``NFW'' \citep{NFW96} dark matter halo. Again we assume that the gas is at the virial temperature of the halo. While a more realistic model would include a clustercentric radius-dependent temperature \citep[e.g.][]{Arnaud2010}, we adopt a single-temperature model to minimize the number of free parameters. Using this model, \citeauthor{Makino1998} predict the core density, core radius, $\beta$-parameter, and X-ray luminosity of clusters as function of halo mass and temperature. The Makino profile is given by 
	\begin{equation}
		\rho(r) = \rho_{\rm 0}~e^{-27b/2}~(1+r/r_s)^{27b/(2r/r_s)} ~.
		\label{eq:gasprofile}
	\end{equation}
	Here $\rho_{\rm 0}$ is the core density of hot gas, while $r_s$ is the usual NFW scale radius. If one assumes the hot gas is at the virial temperature of the halo (Equation~\ref{eqn:Tvir}), the $b$ parameter can be written as
	\begin{equation} 
		b = {2C \over 9\gamma}
		\left[\ln\left(1+{C}\right)- {C \over 1+C}\right]^{-1},
		\label{eq:b}
	\end{equation}  
	with $\gamma$ being a parameter of order unity that determines the efficiency of shock heating; we adopt $\gamma=1.5$. The concentration, $C = R_{\rm vir}/r_s$, was empirically determined by \cite{Bullock2001} to be
	\begin{equation}
		C= \frac{4}{1+z} \left( \frac{M_{\rm vir}}{1.4 \times 10^{14} \rm M_{\odot}} \right)^{-0.13} ~,
		\label{eq:C}
	\end{equation}
	with $z$ the halo's redshift. Finally, the central density of the hot gas, $\rho_{\rm 0}$, can be determined by integrating the Makino density profile out to the halo virial radius and requiring the total mass equal $m_{\rm hot}$:
	\begin{equation}
		\rho_{\rm 0} = 
		\frac{m_{\rm hot}}{4 \pi r_s^3} e^{27b/2}
		\left[ \int_0^C x^2 (1+x)^{27b/2x} dx
		\right]^{-1} ~,
		\label{eq:rho0}
	\end{equation}
	where we have used the change of variable $x\equiv r/r_s$.
	
	We can combine the above Makino profile representations for $b$ (Equation~\ref{eq:b}), $r_s = R_{\rm vir}/C$ (using Equation \ref{eq:C}), and $\rho_{0}$ (Equation~\ref{eq:rho0}) to obtain a $\beta$-model-type expression,
	\begin{equation}
		\label{eq:betafit}
		\rho(r) = {\rho_{\rm 0} A(b)\over 
			[1+(r/r_{\rm c,eff})^2]^{3\beta_{\rm eff}/2}} ~.
	\end{equation}
	The parameters completing the profile are: $A(b)=-0.178b+0.982$, $r_0 \equiv r_{\rm c,eff}= 0.22r_s$ and $\beta_{\rm eff}=0.9b$, all valid over scales $0.01r_s<r<10r_s$.  
	
	In Figure \ref{fig:Densityp} we compare the density profile of \cite{Makino1998} (Equation~\ref{eq:gasprofile}, red solid line) to that of the $\beta$-model
	(Equation~\ref{eq:betaprofile}, blue long-dashed line) as function of radius (in units of $R_{\rm vir}$) for the same total hot gas mass ($m_{\rm hot} = 10^{13} \rm h^{-1} M_{\odot}$). Also over-plotted is the original isothermal profile ($\propto 1/r^2$) used in \textsc{SAGE} (Equation~\ref{eq:isothermal}, gray short-dashed line). The Makino and $\beta$-model profiles agree comfortably well across the entire range plotted, while the isothermal profile is seen to deviate significantly from the other two on scales less than $\sim 0.1 R_{\rm vir}$, where cooling is most significant.

	\subsection{ Jet generation and propagation}
	\label{Sec:RadioJet}
	
	Following the original model described in \cite{Croton2006}, the accretion rate of gas feeding the black hole is approximated by the Bondi-Hoyle formula \citep{Bondi1952},
	\begin{equation}\label{Bondi}
		\dot m_{\rm BH} = \frac{2.5 \pi G^2 m_{\rm BH}^2 \rho_0}{c_s^3} ~,  
		\label{eq:BHaccrOrig}
	\end{equation}
	where $m_\mathrm{BH}$ is the black hole mass and $\rho_0$ is the density of accreting hot gas around the black hole. $c_s \equiv V_{\rm vir}$ and $G$ are the speed of sound in the gas and the gravitational constant, respectively. 
	
	The key unknown above is the central gas density, which can be approximated by assuming a maximal cooling flow in this region, as explained in section~9.1 of \cite{Croton2016}. This leads to an expression for $\rho_0$ that can be inserted back into Equation~\ref{eq:BHaccrOrig}, after which the accretion rate simplifies to
	\begin{equation}\label{eq:BHaccr}
		\dot M_{\rm BH} = \kappa_R {15 \over 16} \pi G~ \mu m_p {kT \over \lambda} M_{\rm BH} ~.    
	\end{equation}
	Equation~\ref{eq:BHaccr} is now written in terms of quantities that \textsc{SAGE} naturally produces. Note that the efficiency parameter, $\kappa_R$, was introduced by \cite{Croton2016} to modulate the effectiveness of the feedback within the cooling--heating cycle.
	
	\begin{figure}
		\centering
		\includegraphics[width=0.5\textwidth]{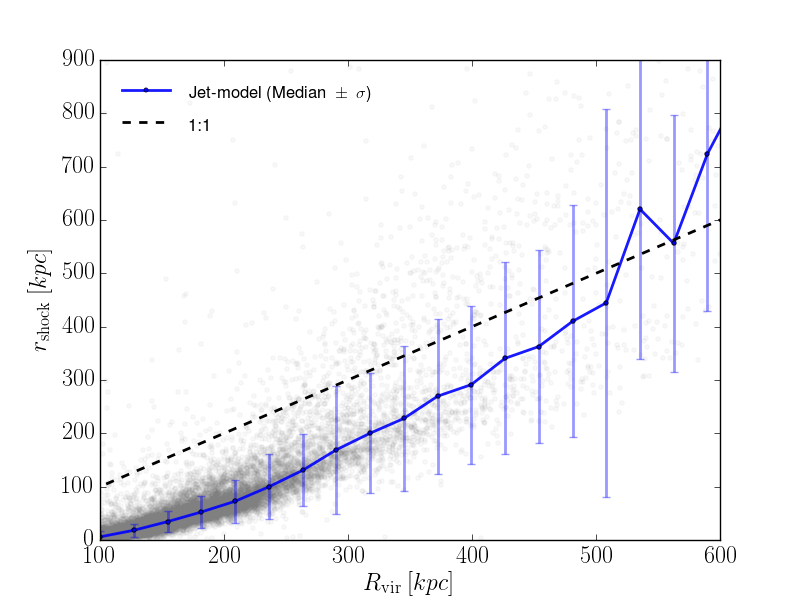}
		\caption{Distribution of shocked radius as function of virial radius. The blue solid line presents the median of the individual model points, with standard deviation given by the error bar. The dashed black line indicates the limiting case when the shock radius extends out to the virial radius.}
		\label{fig:RShock_vir}
	\end{figure}
	\begin{figure}
		\centering
		\includegraphics[width=0.5\textwidth]{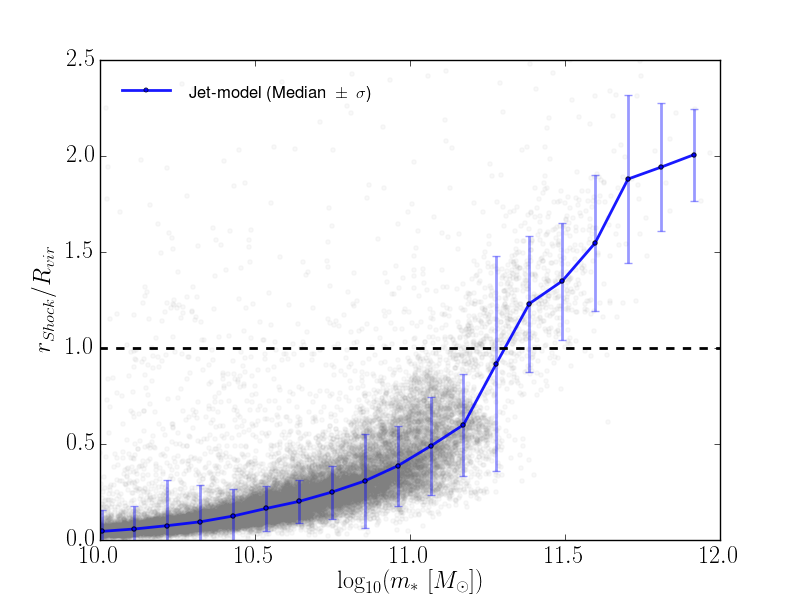}
		\caption{Shocked radius in units of $R_{\rm vir}$ as function of stellar mass, as shown by the gray data points for model central galaxies. The blue solid line presents the median of the individual model points, with standard deviation given by the error bar. The dashed black line indicates the limiting case when the shock radius extends out to the virial radius.}
		\label{fig:Rshock_hist}
	\end{figure}
	
	Moving beyond this original accretion implementation, we look to build a physical model of the outflows produced as a result of gas accretion onto a black hole. There are two different accretion states that can both happen when removing angular momentum from the accretion disk by viscosity
	\citep{Narayan1998,Meier2001,Koerding2006,Fender2004}. The ``hot mode'' occurs for low accretion rates with respect to the Eddington rate 
	($\dot{M}_{\rm BH} < \alpha_{\rm crit} \dot{M}_{\rm edd}$). Here, the flow is geometrically thick and optically thin, which results in longer cooling times and more powerful jets (for a given black hole mass and accretion rate). The thermal energy of the inflowing gas is advected inward, known as an advection dominated accretion flow \citep[ADAF;][]{Narayan1995}.

We relate ``hot mode" jet power to the accretion rate via the following relation:
	\begin{equation}\label{eqn:Qjet_eta}
		Q_{\rm jet,hot} = \eta \dot{M}_{\rm BH} c^2 ~,
	\end{equation}
	where $c$ is the speed of light and $\eta$ is the jet efficiency, estimated using relativistic magnetohydrodynamical simulations of jet generation to have  typical values between 0.002 and 1;  the exact value is largely set by black hole spin \citep{Benson2009}. In Section \ref{sec:RadioAGN} we use observations to constrain the best-fitting value of $\eta$ to 0.35.
	
	In contrast, the ``cold mode'' of AGN accretion occurs when inflow rates are high compared with the Eddington rate ($\dot{M}_{\rm BH} > \alpha_{\rm crit} \dot{M}_{\rm edd}$). Here, the accretion flow can be described by the standard thin disk solution of \cite{Shakura1973}, where the disk is geometrically thin and optically thick. This produces outflows that are radiatively efficient and hence the AGN has weaker jets (for a given black hole mass and accretion rate) in comparison to the ADAF case (Narayan et  al., 2002). 
	The cold mode jet power is given by \citep[][ Equation 5]{Shakura1973,Meier2001}:
	\begin{equation}
		Q_{\rm jet,cold} = 6.3 \times 10^{-5}\left( \frac{M_{\rm BH}}{10^9 \rm M_{\odot}\/} \right)^{-0.1}  \dot{m}_{\rm BH} ^{0.2}~ (\dot{M}_{\rm BH} c^2 ) [W] ~,
		\label{eqn:QjetTD}
	\end{equation}
where  $\dot{m}_{\rm BH}\equiv \frac{\dot{M}_{\rm BH}}{\dot{M}_{\rm edd}}$.

Note that the parameter $\alpha_{\rm crit}$ defines the transition from hot to cold accretion in units of the Eddington rate and is set at 0.03 throughout this work \citep[see also][]{Shabala2009,Merloni2008}. In our model we generate both ADAF and thin disk jets, with different jet generation efficiencies.

\subsubsection{Radio source expansion}
\label{Sec:RadioExpan}

The accretion disk outflows just described drive jets into the surrounding gas, creating expanding cavities and producing synchrotron emission seen as radio lobes. The point at which the jet impacts the far side of the cocoon is seen observationally as a hotspot. The plasma swept up from the jet injection causes these radio lobes to expand supersonically; this characterises the cocoon expansion, the evolution of which we now quantify.

Using the gas density profile described in Section~\ref{sec:Densityprofile}, and assuming a cocoon axial ratio $R_T$ for a jet with kinetic power $Q_{\rm jet} / 2$, the dynamical model of \cite{Kaiser1997} characterises the evolution of the cocoon radius as
\begin{equation}\label{eqn:Rcocoon}
	r_{\rm cocoon} (t) = a_D r_0 \left( \frac{t}{\tau} \right)^{3/(5-\beta)} ~.	
\end{equation}
Here time, $t$, is normalised to a characteristic time-scale, $\tau = \left( \frac{r_0^5 \rho_0}{Q_{\rm jet}} \right)^{1/3}$, while the dimensionless constant, $a_D$, follows from the work of \cite{Kaiser1997} and \cite{Shabala2009}:
\begin{equation}\label{eqn:a_D}
	a_D = \left[ \frac {R_T^4\ (\Gamma_{\rm x}+1) (\Gamma_{\rm l}-1) (5-\beta)^3} {18 \pi \left( 9 \left[ \Gamma_{\rm l} + (\Gamma_{\rm l}-1) R_{\rm T}^2/2 \right] -4-\beta \right)} \right]^{1/(5-\beta)} ~. \nonumber
\end{equation}
 Assuming a power law density profile of $\rho = \rho_0 (r/r_0)^{-\beta}$ for the cocoon expansion 
	where $0< \beta < 2$, and axial ratio between $1.3 < R_T < 6$, the value for $a_D$ is
	in the range of 0.8-1.8 (see Table \ref{tab:value}). We adopt $a_D=1$, a value typical for radio sources in the inner regions of clusters.

\begin{figure}
	\centering
	\includegraphics[width=0.5\textwidth]{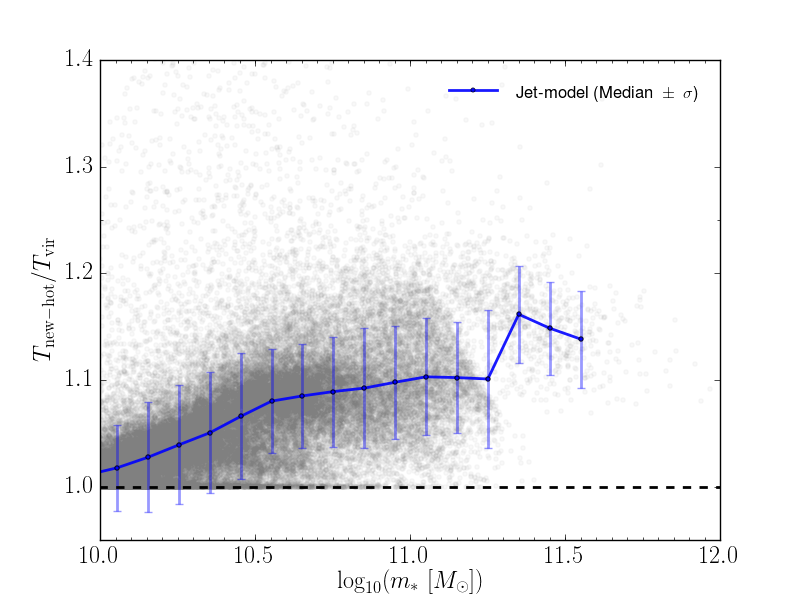}
	\caption{New hot gas temperature in units of $T_{\rm vir}$ as a function of stellar mass. Results are show for central galaxies only, with the blue line and error bars indicating the median and standard deviation of the distribution in each stellar mass bin, respectively. The dashed black line indicates where the temperature remains at the virial value.}
	\label{fig:Temp_hist}
\end{figure}

\subsubsection{The cocoon shock radius}
\label{Sec:Rshock}

Following \cite{Kaiser1997}, the relation between the maximum radius of the shocked gas and the radius of radio emitting cocoon plasma is given by
\begin{equation}\label{eqn:Rshock}
	r_{\rm shock} = {1 \over \lambda} r_{\rm cocoon} ~,
\end{equation} 
where the maximum radius is calculated using the AGN active time ($t_{on}$, describe below in Section \ref{sec:ittermit}) and the cocoon radius (Equation \ref{eqn:Rcocoon}). $\lambda$ is dependent on the density profile, and we have assumed a self-similar solution for the shape of the cocoon,
\begin{equation}\label{eqn:lamdashock}
	1 - \lambda^3 = {15 \over 4(11-\beta)} ~.
\end{equation} 
 
 Observations show the existence of somewhat rare, extremely powerful AGN outbursts in massive galaxies \citep{Rawlings2004,Miley2008,Shabala2011,Fabian2012}. Their size extends many hundreds of kpc, and their life-times can be up to $10^8$ yr \citep{Shabala2008}. Radio sources sizes, denoted $D$, have a typical range of between 1 and $\sim$1000 kpc \citep{Willott1999,Sadler2007}.  In our model the shocked gas expands up to $\sim 900$ kpc, as can be seen in Figure \ref{fig:RShock_vir}, approximately consistent with such observations. Similarly, in Figure \ref{fig:Rshock_hist} we show the distribution of shock radii as function of stellar mass. Together, these figures indicate that AGN feedback can clear hot and cooling gas out to large radii. In our model, larger radio sources should preferentially be found in more dense environments. On the one hand, denser environments imply slower expansion speeds (through Equation \ref{eqn:Rcocoon}); on the other hand, more massive black holes found in such environments have both higher jet powers (Equations \ref{eqn:Qjet_eta}, \ref{eqn:QjetTD}) and are longer lived (Equation \ref{eq:t_on}). As shown in Figure \ref{fig:RShock_vir}, the net effect is for larger radio sources to preferentially inhabit dense environments in our model. Observationally, many large double-lobed radio galaxies are found in poor environments; on the other hand, many small-scale jets are also associated with Seyfert galaxies \citep[e.g. ][]{Ulvestad1984,Kaviraj2015}. The apparent compactness of many radio AGN may at least in part be due to the lack of sensitivity to diffuse low-surface brightness features in FR-I radio galaxies \citep{Shabala2017}. Our simple model cannot address this selection effect, which is deferred to future work.
Given the simplicity of our model we resist the temptation to make more quantitative predictions than this, but it is an area for future work.  
The feedback is done in two ways, through heating of the gas, and uplifting of the gas, which we describe in the following sections.

\subsubsection{Gas heating and cooling}\label{Sec:GasNewHot}

According to \cite{Alexander2002}, and following the study by \cite{Shabala2009}, the difference between the mean isothermal temperature in the shocked gas and the post-shock temperature can be described by
\begin{equation}\label{eqn:Tshock}
	T_{\rm shock} = {15 \over 16} {3-\beta \over 11- \beta} \left({\mu m_H \over k_B} \right) \dot{r}_{\rm shock}^2 ~,
\end{equation}
where the shock velocity is given by $\dot{r}_{\rm shock} = {3 \over  5-\beta} {r_{\rm shock} \over t_{\rm on}}$, where $t_{\rm on}$ is the active time of the AGN and is defined below in Section \ref{sec:ittermit}.
We can then obtain the following integrated hot gas mass using the \cite{Makino1998} profile for a spherical shocked mass distribution, given by
\begin{equation}\label{eqn:Mshock}
	m_{\rm shock} (r_{\rm shock}) = 16 \pi \rho_{g0} r_0^3 \left[\ln \left(1+ {r_{\rm shock} \over r_0} \right)
	- {r_{\rm shock} \over r_{\rm shock} + r_0}\right] ~.
\end{equation}
The shocked mass can never exceed that of the mass of hot gas. 
When the gas is shock heated we assume an instantaneous mixing of this gas with the hot gas of the halo, resulting in a new hot halo temperature of
\begin{equation}\label{eqn:Tnewhot}
	T_{\rm new-hot} = T_{\rm hot} + {m_{\rm shock} \over m_{\rm hot}} (T_{\rm shock}-T_{\rm hot})  ~.
\end{equation}

Figure \ref{fig:Temp_hist} shows the distribution of this new hot gas temperature in units of the original virial temperature, given as a function of stellar mass. In general, AGN in our model act to raise the temperature of the gas by between a few to $20 \%$, depending on the stellar mass of the central galaxy in the halo. In this model, we only change the temperature if $T_{\rm shock} > T_{\rm vir}$.

Hot halo gas is not static but cools over time. The cooling time of this gas at each radius is taken as the ratio of its specific thermal energy to the cooling rate per unit volume, 
\begin{equation}
	t_{\rm cool}(r) = \frac{2}{3} \frac{\mu m_p k_B T_{\rm new-hot}}{\rho(r) \Lambda(T_{\rm new-hot},Z)} ~,
\end{equation}
where $\Lambda(T,Z)$ is the cooling function and depends on the temperature, $T$, and metallicity, $Z$. 
In such models a cooling radius $r_{\rm cool}$ is then defined as the radius at which $t_{\rm cool}$ is equal to a characteristic time-scale of the system, such as its age, time since last major merger, or dynamical time. Following the standard implementation of SAGE we take the last of these.

For the so-called \textit{hot accretion mode}, when $r_{\rm cool} < R_{\rm vir}$, the cooling rate of the hot gas is calculated using
\begin{equation}
	\dot{m}_{\rm cool} = 4 \pi \rho_{g}(r_{\rm cool}) r_{\rm cool}^2 \dot{r}_{\rm cool} ~,
	\label{coolingrate_1}
\end{equation}
where $\rho_{g}(r_{\rm cool})$ is the \cite{Makino1998} profile gas density at the cooling radius. With this, the cooling rate can be rewritten as  
	\begin{eqnarray}
	\dot{m}_{\rm cool} = 4 \pi \rho_{\rm 0} \left(r_{\rm cool}^3 \over t_{cool}\right)  \left({r_{\rm cool} \over r_s}\right) \left({e^{13.5b} \over 13.5b}\right) 	
	 \nonumber \\
\times  \left({\left(1+{r_{\rm cool} \over r_s}\right)^{13.5b\left({r_s \over r_{\rm cool}}\right)} \over \ln \left(1+ {r_{\rm cool} \over r_s}\right) - {r_{\rm cool} \over r_s +r_{\rm cool}}}\right).
	\label{coolingrate}
\end{eqnarray}

Alternatively, in the so-called \textit{cold accretion mode}, when $r_{\rm cool} > R_{\rm vir}$, the hot halo never forms and any infalling gas captured in the dark matter halo potential falls towards the centre on a free fall time-scale, $R_{\rm vir}/V_{\rm vir}$, given by 
\begin{equation}
	\dot{m}_{\rm cool} = \frac{m_{\rm hot}}{R_{\rm vir}/V_{\rm vir}} ~.
\end{equation}

\begin{figure}
	\centering
	\includegraphics[width=0.5\textwidth]{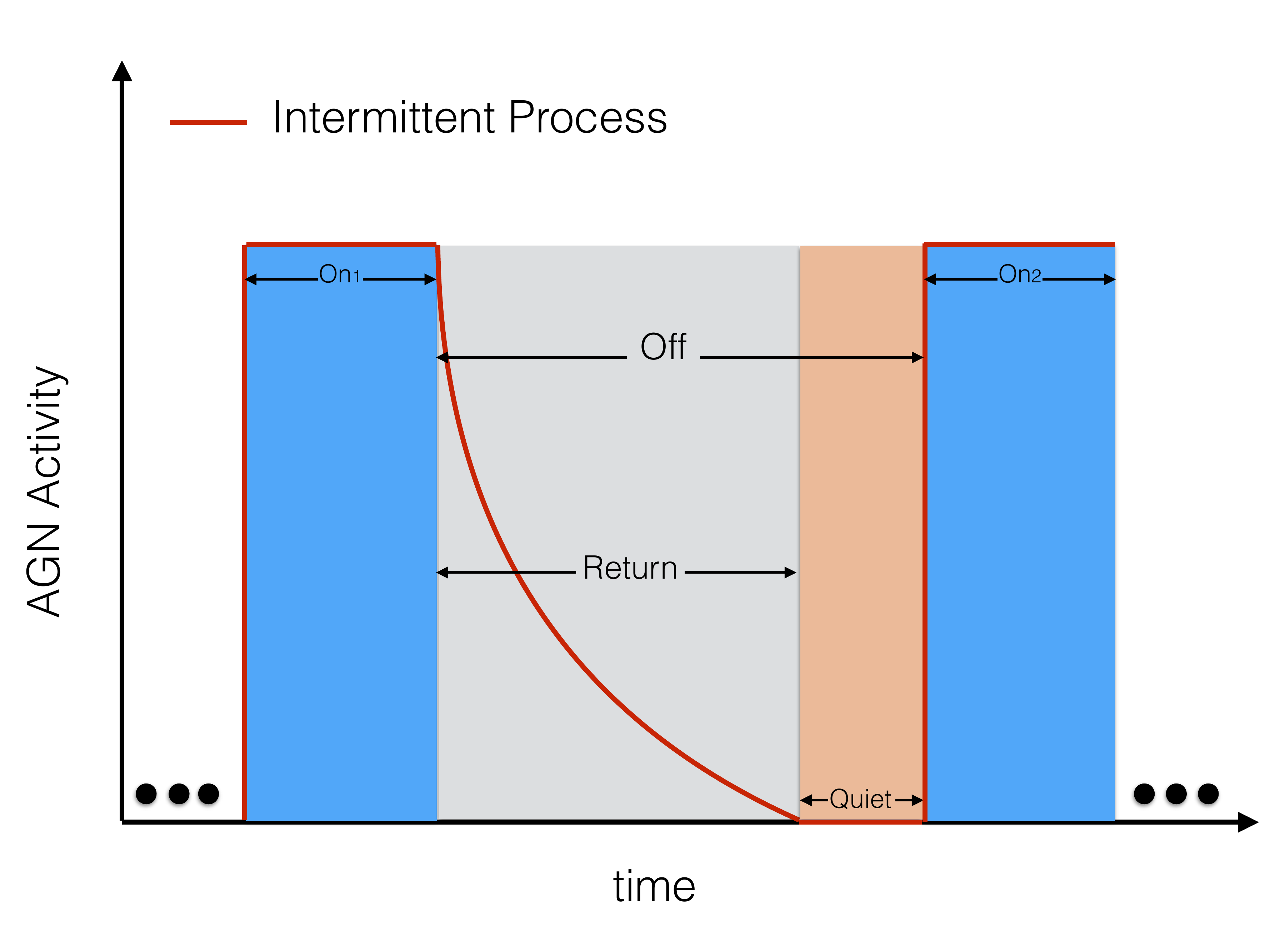}
	\caption{A schematic of our intermittent physically motivated model for AGN feedback. The red line shows the time evolution of the jet. During the ``on" and ``return" phases only gas that has not been overrun by the shock can cool. During the ``quiet" phase cooling is allowed to proceed as usual (Equation \ref{eq:F_cool}).
	}
	\label{fig:Cartoon_itter}
\end{figure}

\subsubsection{The AGN duty cycle and its effect on cooling}
\label{sec:ittermit}

Left unchecked, massive dark matter halos in particular can collect prodigious amounts of hot gas which will lead to runaway cooling at rates that are unsupported by the observations. Our AGN model provides an energy counterbalance to such cooling through the heating resulting from an AGN jet. Over long time-scales (100's of Myr to many Gyr), such energy injection can be approximated as uniform and constant, as assumed in \cite{Croton2006}. However, a more realistic model will attempt to describe the intermittent nature of black hole accretion and their resulting outflows and properties.

To this end, suppose that we have a black hole of a given mass being fed gas from its surrounding medium through an accretion disk. We simplify the otherwise complicated resulting AGN duty cycle by breaking it into four primary parts: the time that the jet is on, $t_{\rm on}$, inflating the cocoon as described above; the switching off of the AGN and hence jet when accretion stops; the time it takes for the cocoon to disappear once the jet pressure has been removed, $t_{\rm return}$; and finally the subsequent period of quiescence after the cocoon has dissipated but before the next episode of accretion reignites the AGN and jet, $t_{\rm quiet}$. This cycle is illustrated in Figure~\ref{fig:Cartoon_itter}, and we now look to quantify each phase using the previously established relations. 

Observationally, the AGN fraction as a function of stellar mass was used by \cite{Turner2015} to estimate an expression for $t_{\rm on}$, approximately given by
\begin{equation}\label{eq:t_on}
	t_{\rm on} = 120 ~ \left[{m_* \over 10^{11} \rm M_{\odot}}\right]^{0.7} ~ \rm Myr ~.
\end{equation} 
Characteristic time-scales for $t_{\rm on}$ range from between $2 \times 10^{7}$ to $5 \times 10^{8}$ years for stellar masses between $10^{10}$ and $10^{12}\rm M_{\odot}$. For very low mass galaxies this expression implies radio sources that are so small as to do no feedback.

Once support from the jet for the cocoon has been removed the time for it to collapse back down can be estimated using the free fall time from the shock radius, which we parametrise as 
\begin{equation}\label{eq:t_returne}
	t_{\rm return} = 2 {r_{\rm shock} \over c_s} \approx 2 {r_{\rm shock} \over V_{\rm vir}} ~.
\end{equation} 
Here, $c_s$ is the sound speed in the gas which we approximate using the virial velocity of the parent halo, $V_{\rm vir}$. The factor of two is an empirically chosen value we use to prevent overcooling in our model. Physically, the slow (slower than the sound speed) return of the shocked gas is due to processes we do not model here, such as the movement of buoyant bubbles once the jet switches off, and the slow mixing of the heated gas throughout the cluster \citep{Basson2003, Soker2015, Yang2016}.

The total time the AGN is off, $t_{\rm off}$, can be calculated from $t_{\rm on}$ using the observed duty cycle of AGN in the local Universe. \cite{Best2005} and \cite{Shabala2008} both measured the radio loud AGN fraction as a function of galaxy mass; using Figure~3 of \cite{Best2005} the duty cycle can crudely be approximated as
\begin{equation}
	\delta= 0.05 \left[{m_* \over 10^{11}\rm M_{\odot}}\right]^{1.5} ~,
	\label{eq:Duty}
\end{equation}
from which $t_{\rm off}$ is simply
\begin{equation}\label{eq:t_off}
	t_{\rm off} (m_*, t_{\rm on}) = t_{\rm on} \left[{1 \over \delta} -1\right] ~.
\end{equation}
It then follows that $t_{\rm quiet}$ is just the difference between the total AGN off time and the cocoon return time,
\begin{equation} \label{eq:t_static}
	t_{\rm quiet} = t_{\rm off} - t_{\rm return} ~.
\end{equation}

The presence of inflated cocoons from the AGN is expected to modify the cooling rate in the hot halo in multiple ways. 
For our model we make a simple approximation and suppress cooling whenever a cocoon is present, with the degree of suppression proportional to the hot gas mass fraction between the shock radius and cooling radius. In other words, only gas that has not been overrun by the shock can cool.
This approximation is reasonable, as the active phase corresponds to at most a third of the total cycle time (see Equation~\ref{eq:Duty}), and significantly less for lower mass galaxies.
Furthermore, jet driven bubbles uplift the shocked gas further away from the galaxy centre, thus stopping it from cooling. Once the cocoon expansion stops, gas can re-fill the evacuated cavity. If this ``return" time is less than the ``off" time the system will experience a period of quiescence and cooling will be restored. Across the entire cycle the cooling rate may be further modified through the altered hot gas temperature (see Figure \ref{fig:Temp_hist}).

In summary, the time-averaged fraction of cooling in the presence of the AGN duty cycle is
\begin{equation}\label{eq:F_cool}
	f_{\rm cool} = {t_{\rm quiet} \over t_{\rm on} + t_{\rm off}} + \left[{t_{\rm on}+t_{\rm return} \over t_{\rm on}+t_{\rm off}}\right] \left({m(r_{\rm cool})-m(r_{\rm shock}) \over m(r_{\rm cool})}\right)~.
\end{equation} 
We include only the first term in the case of modified cooling due to jet heating that leads to higher hot gas temperatures, only the second term in the case of feedback and altered cooling from gas uplifting, and both terms in the full model.
The new cooling rate is given by
\begin{equation}\label{eq:t_static}
    \dot{m'}_{\rm cool} =  f_{\rm cool} \, \dot{m}_{\rm cool} ~.
\end{equation}

\begin{figure}
	\centering
	\includegraphics[width=0.5\textwidth]{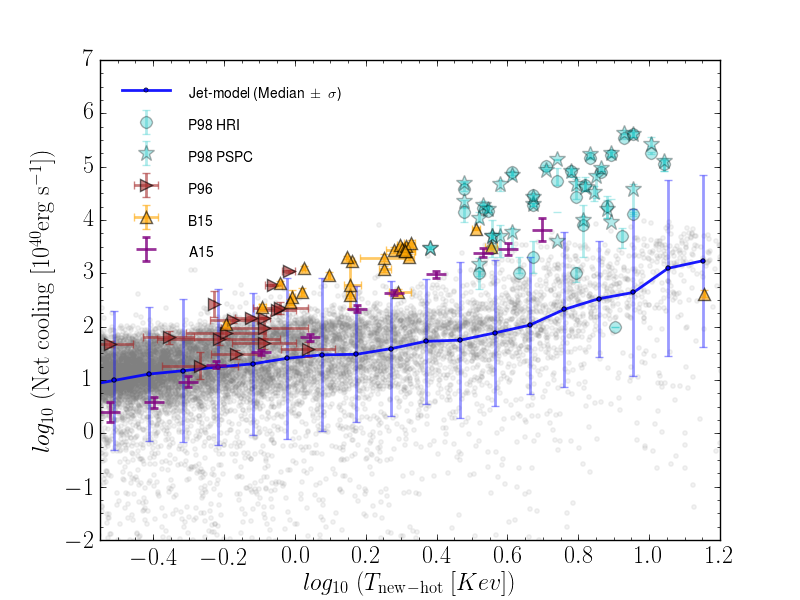}
	\caption{Net cooling rates as function of hot halo gas temperature at z = 0. Gray data point show model central galaxies, with the blue line and error bars indicating their median plus standard deviation. Circled and starred points with error bars show the observational data of galaxy clusters from \citet[P98]{Peres1998}, measured with the High Resolution Imager (HRI) and Position Sensitive Proportional Counter (PSPC) on ROSAT, respectively. Triangular points with errors show the observational data of galaxy groups from \citet[P96]{Ponman1996} and \citet[B15]{Bharadwaj2015}. Stacked X-ray observations surrounding local brightest cluster galaxies from \citet[A15]{Anderson2015} are shown with the horizontal lines with errors.}
	\label{fig:Cooling_Temp}
\end{figure}

\begin{figure}
	\centering
	\includegraphics[width=0.5\textwidth]{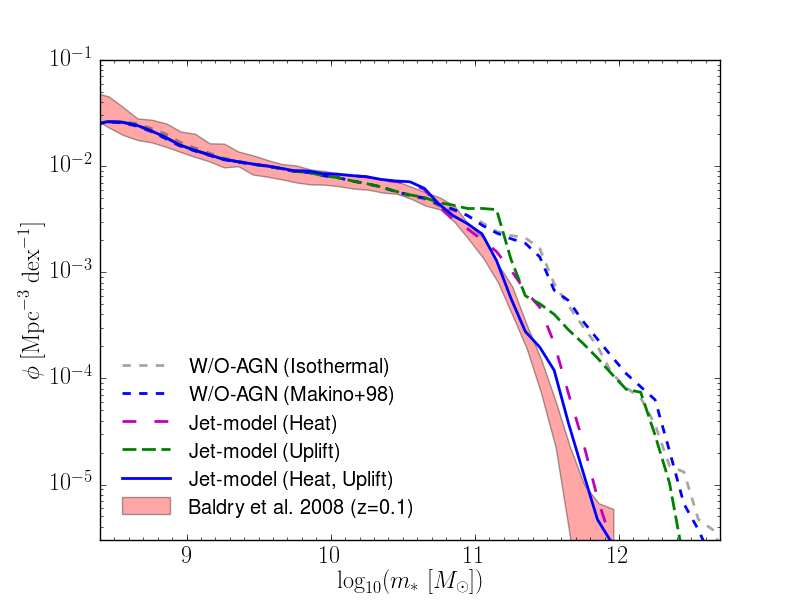} 	
	\caption{The stellar mass function of model galaxies for a range of different AGN assumptions. We highlight the local observed function from the SDSS using the red shaded region, as measured in \citet{Baldry2008}. The original isothermal and \citet{Makino1998} density profiles with all AGN switched off are given by the gray dashed line and blue dashed line, respectively. The effect of using the new hot gas temperature (`heat') and AGN duty cycle (`uplift')  are shown with the red and green long-dashed lines, respectively. The `heat' and gas `uplift' (see Section~\ref{sec:ittermit}) prescriptions together is given by the blue solid line, our final model.}    	
	\label{fig:Massfunc}
\end{figure}

\begin{figure}
	\centering
	\includegraphics[width=0.5\textwidth]{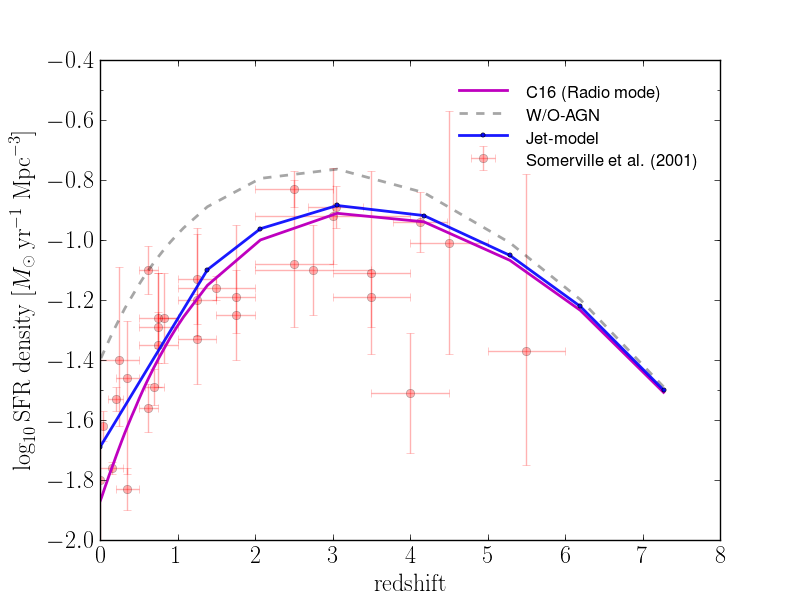}	
	\caption{The average star formation rate density history of the Universe produced using our jet model (blue line), compared to observational compilation from Somerville et al. (2001) (red points), the \citet{Croton2016} model (magenta long-dashed line), and our model with AGN feedback switched off (gray short-dashed line).}
	\label{fig:SFRD}
\end{figure}

\section{Results}

\subsection{Model constraints}\label{Sec:Constrains}

Our AGN jet model makes three important changes to the previous SAGE radio-mode prescription. Namely, it uses a more realistic hot gas density profile to calculate cooling, the AGN jet can heat the hot gas to higher temperatures than before, and the combined effect of this new temperature and the jet inflated cavities act to alter the cooling rate in a more realistic way. To calibrate this model we use a set of observables that the output must reasonably compare with, and in a physically sensible way. Only then can we extend our analysis to explore predictions and consequences that can be compared with future observations.

To start, in Figure~\ref{fig:Cooling_Temp} we show the cooling rates from our model at z=0 as function of their hot halo gas temperature (Equation~\ref{eqn:Tnewhot}). We compare with a number of X-ray observations of gas surrounding local galaxy clusters and groups: \citet{Ponman1996} and \citet{Bharadwaj2015} who evaluate the the bolometric luminosity within $r_{500}$; \citet{Peres1998} who measure the luminosity inside the cooling radius, and \citet{Anderson2015} who use stacked observations of X-ray emission around locally brightest cluster galaxies. This range of observations serve as a good general indicator of how our new model compares. As the gray model data points and solid median line reveal, our cooling--temperature relation provides a reasonably good fit to the observations, except perhaps for the hottest cluster halos, where we tend to under-predict cooling somewhat. The jet model fares considerably better than the original \citet{Croton2006} and \citet{Croton2016} models, which over--and under--predict these observations by a significant amount, respectively \citep[see e.g. their figure 7]{Croton2016}.

The consequence of cooling is a pooling of cold gas in the galactic disk which then leads to star formation \citep[see e.g. ][]{Croton2006}. A key motivator to include AGN feedback in semi-analytic models is to understand how the suppression of this infalling gas then leads to a suppression of the formation of new stars. The effectiveness of this feedback, and at the right mass scale, is usually quantified using the galaxy stellar mass function, the shape of which is moulded by both AGN and supernovae. As in previous works, we use the high mass end and knee of the local stellar mass function to constrain our jet model (the low mass end is constrained by other feedback processes and not the focus of the present work).

Figure~\ref{fig:Massfunc} shows the effect of the jet model on the formation of high mass galaxies compared to the SDSS observations of \cite{Baldry2008}.  First we present the stellar mass function for the original SAGE model without any AGN feedback (gray dashed line), then systematically add the changes described in the previous sections until we arrive at our default version. This version includes the updated hot gas density profile (blue short-dashed line), and both gas heating by the AGN jet (red long-dashed line) and also mass transport through uplifting (blue solid line, our final model).

Lastly, we consider the star formation rate density evolution of the galaxy population with time. In Figure~\ref{fig:SFRD} we plot the final model by the blue line and two bounding cases: the original SAGE model (magenta long-dashed line) and our model but with AGN feedback turned off (gray short-dashed line). As can be seen, both SAGE and our new jet model matches the observational compilation of \cite{Somerville2001} reasonably well. This is in contrast to when AGN feedback is turned off, which overproduces star formation at all redshifts except the highest.

\begin{figure}
	\centering
	\includegraphics[width=0.54\textwidth]{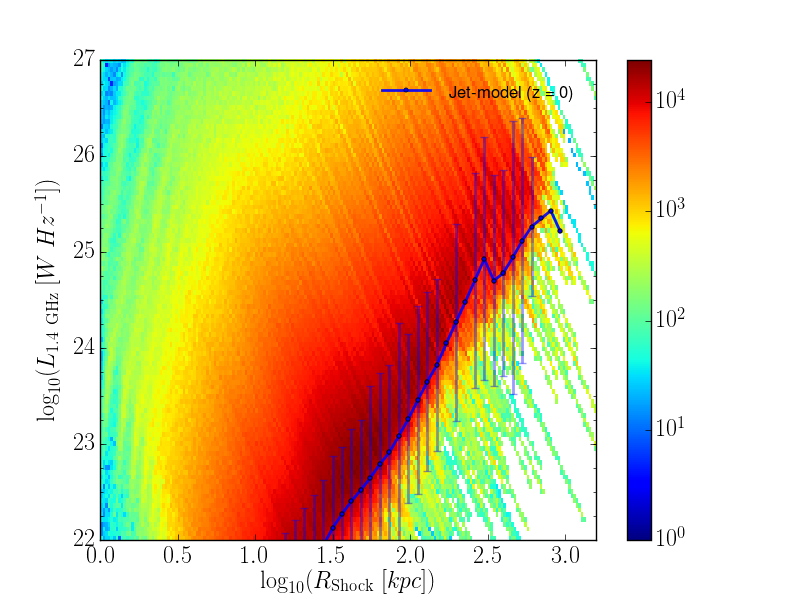}
	\caption{The distribution of radio luminosity as function of shock radius weighted by AGN active time, $t_{on}$, and color coded by the abundance of size-luminosity pairs in each bin size. The blue line and error bars indicate the median and standard deviation of the central galaxies in the model at z = 0, respectively.}  
	\label{fig:Lradio_Rsh}
\end{figure}

\begin{figure}
	\centering
	\includegraphics[width=0.49\textwidth]{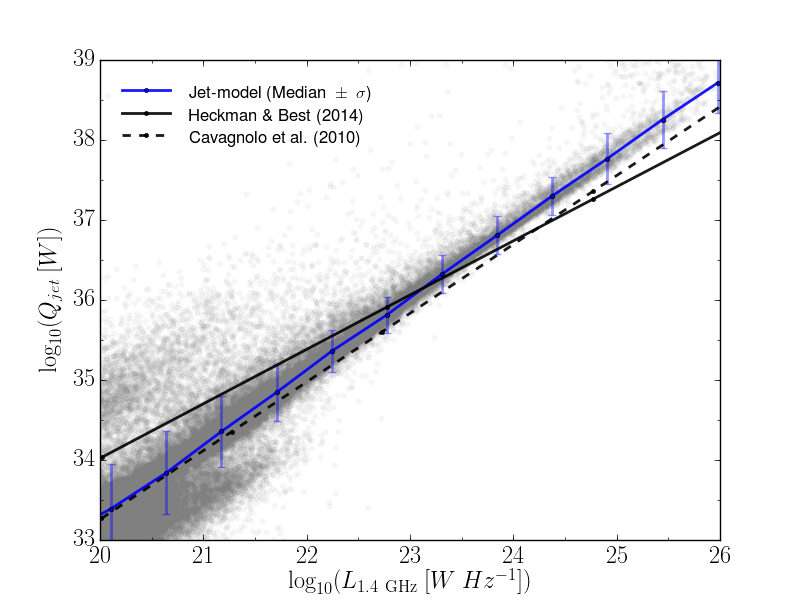}
	\caption{The model relationship between derived 1.4 GHz radio luminosity and jet power, $Q_{\rm jet}$. Gray data points show the prediction for central galaxies in the model, while the blue line and error bars show the median and standard deviation of the points, respectively. This is compared to the observed cavity relation derived by  \citet{Cavagnolo2010} (black dashed line) and  \citet{Heckman2014} (black solid line).
	}
	\label{fig:Lradio-Qjet}
\end{figure}

\subsection{Deriving radio luminosities}
\label{Sec:JetProper}

We now consider the AGN properties of our model galaxies, a new feature of this work, with a focus on comparing with observables. In particular, we look at the cocoon shock size, jet power, and 1.4GHz radio luminosity.

In our model the jet is assumed to expand into a hot atmosphere with a Makino density profile, characterised by core radius $r_0$, core density $\rho_0$, and power-law outer slope $\beta_{\rm eff}$ (see Section~\ref{sec:Densityprofile}). The resulting radio lobes can be seen by their synchrotron emission. This depends on the cocoon magnetic field energy density, $u_B$, the distribution of Lorentz factors $\gamma$ of the emitting electrons, $n(\gamma)$, and cocoon volume, $V$. 
The energy due to the emission of synchrotron radiation is mostly injected at the hotspot between a minimum and maximum Lorentz factor, $\gamma_1$ and $\gamma_2$, with the initial energy distribution given by
\begin{equation}\label{eqn:n_gama}
n(\gamma_i) =  k_e  \gamma_i^{-p} ~.
\end{equation}
The electrons inflate the radio-emitting cocoon and lose energy through adiabatic expansion, synchrotron emission, and inverse Compton upscattering of Cosmic Microwave Background photons.

Following \cite{Kaiser2007} and \cite{Shabala2013}, under the assumption of equipartition between particle and magnetic field energy density the cocoon radio luminosity at frequency $\nu$ can be written
\begin{equation} \label{eqn:Lradio_u_B}
L_{\rm \nu} = A_1 u_B^{\frac{5+p}{4}} V \nu^{\frac{1-p}{2}}  (1+z)^{\frac{(3-p)}{2}} ~,
\end{equation}
where $p$  is the power-law index of the electron energy distribution, and $\alpha=(1-p)/2$. Using the definition of the spectral index for synchrotron emission, $F_\nu \propto \nu^{\alpha}$, a reasonable hotspot value for radio galaxies with $\alpha$ between $-0.55$ and $-0.85$ is $p=2.1-2.7$. 
\begin{figure}
	\centering
	\includegraphics[width=0.49\textwidth]{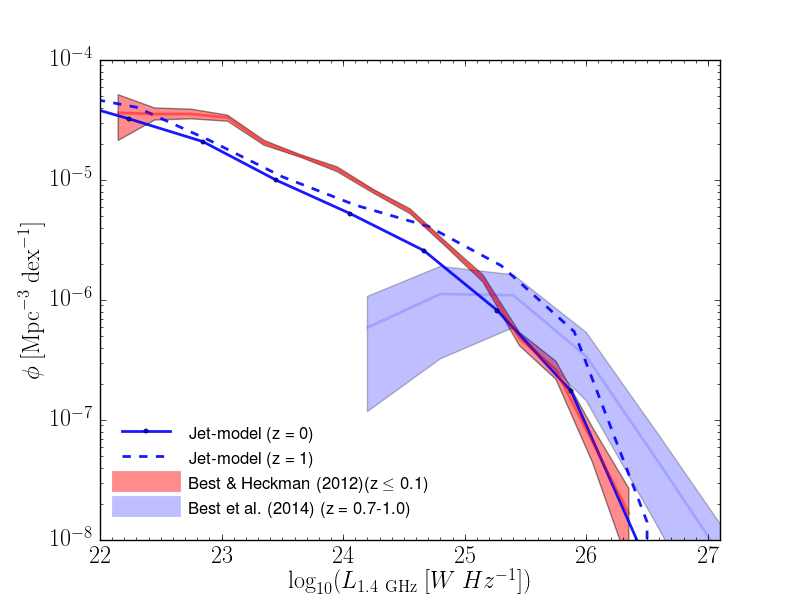}
	\caption{The model radio luminosity function in two redshift bins of z=0 and z= 1. We compare to the observational results of \citet{Best2012} in the local volume (red shading), and 0.7-1.0 (blue shading).}
	\label{fig:RLF}
\end{figure} 
$A_1$ is a numerical constant given by
\begin{equation} \label{eqn:A_1}
A_1 = \frac{16 \pi^2 r_e}{c} \left( \frac{q}{m_e} \right)^{\frac{(p+1)}{2}} \left( 2 \mu_0 \right)^{\frac{(p+1)}{4}} \frac{C_2(p)}{f(p, \gamma_1, \gamma_2)} ~, 
\end{equation}
where $r_e$ is the classical electron radius, $q$ and $m_e$ are the electron charge and mass respectively, $\mu_0$ is the permeability of free space, and $f(p, \gamma_1, \gamma_2) = \int_{\gamma_1}^{\gamma_2} \gamma^{1-p} d\gamma$. The function $C_2(p)$ can be expressed as
\begin{equation}\label{eqn:C_2P}
C_2(p) = \frac{3^{p/2}    }{2^{\frac{p+13}{2}} \pi^{\frac{p+2}{2}}} \frac{\Gamma_{\rm fn} \left( \frac{p+1}{4}  \right) \Gamma_{\rm fn} \left( \frac{p}{4} + \frac{19}{12}  \right) \Gamma_{\rm fn} \left( \frac{p}{4} - \frac{1}{12} \right) }{\Gamma_{\rm fn} \left( \frac{p+7}{4}  \right)} \nonumber ~.
\end{equation} 
Here, $\Gamma_{\rm fn}(z) = \int_0^\infty t^{z-1} e^{-t} dt$ is the Gamma function \citep{Worrall2009}, and for $p = 2.6$, $C_2(p) = 2.04 \times 10^{-3}$. The results explored in this paper assume the best values described in Table~\ref{tab:value}.

For self-similar expansion, the cocoon volume is $V = \pi R_{\rm T}^2 D^3$, where $R_{\rm T}$ is the axial ratio of the source taking values between 1.3 and 6 \citep{Leahy1984,Leahy1989}, and $D = 2 r_{\rm shock}$ is the diameter of the cocoon. It can be shown that the magnetic field energy density, which is proportional to the cocoon pressure, can be written \citep{Kaiser1997,Kaiser2007,Kaiser2008,Shabala2013} as
\begin{equation} \label{eqn:u_B}
u_B = \frac{r f_p}{(r+1)(\Gamma_l -1)} \left( \rho_0 r_0^\beta \left(\frac{Q_{\rm jet}}{2}\right)^2 \right)^{1/3} (2 r_{\rm shock})^{-(4+\beta)/3} ~, 
\end{equation}
where $r=\frac{p+1}{4}$, and the constant $f_p$ is taken from \cite{Kaiser2007},
\begin{eqnarray}\label{eqn:f_p}
f_p &=& \frac{18 a_D^{2(5-\beta)/3}}{(\Gamma_{\rm x} +1)(5 - \beta)^2 R_{\rm T}^2} ~.	
\end{eqnarray} 
The radio luminosity is then 
\begin{eqnarray}
L_{\rm \nu} &=& A_1 \pi R_{\rm T}^2 \left(  \frac{r f_p}{(r+1)(\Gamma_l -1)}  \right)^{\frac{(5+p)}{4}} \nu^{(1-p)/2} \nonumber \\ 
&\times& \left( 1+z \right)^{\frac{3-p}{2}} \left( \rho_0 r_0^\beta  \left(\frac{Q_{\rm jet}}{2}\right)^2 \right)^{\frac{5+p}{12}} \nonumber \\ 
&\times& (r_{\rm shock})^{3-\left( \frac{4+\beta}{3} \right) \left( \frac{5+p}{4} \right)}  ~,
\label{eqn:Lradio}
\end{eqnarray}
Here, $Q_{\rm jet}$ is the total kinetic power from both radio jets.
\begin{figure}
	\centering
	\includegraphics[width=0.5\textwidth]{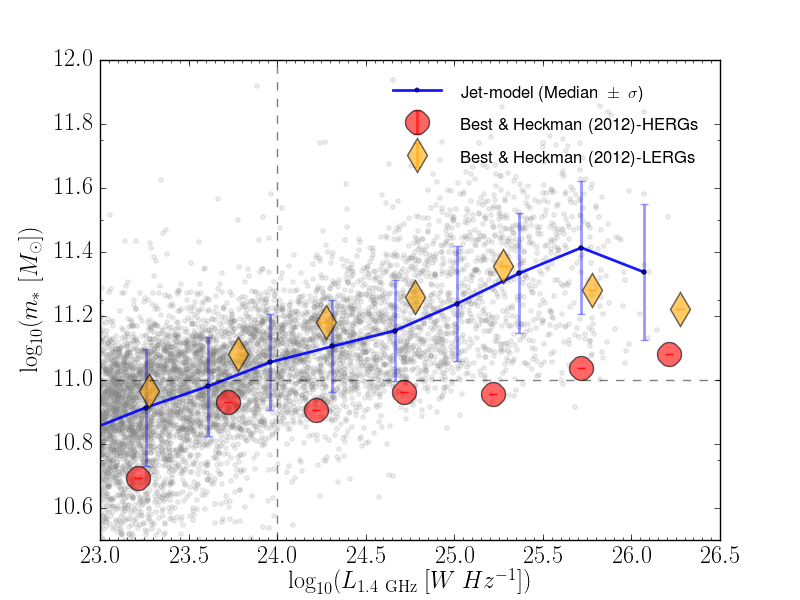}
	\caption{The model stellar mass-radio luminosity relation distribution, compared with the observations of HERGs and LERGs from \citet{Best2012}. Gray data points show the prediction for central galaxies in the model (which we expect to be LERGs), while the blue line and error bars indicate the median and standard deviation of the points, respectively. Dashed lines show the approximate completeness limits of the \citet{Best2012} sample.} 
	\label{fig:Lradio_mass}
\end{figure}

For a given $t_{\rm on}$, higher jet powers yield both larger sizes (Equation \ref{eqn:Rcocoon} and Equation \ref{eqn:Rshock}) and radio luminosity (Equation \ref{eqn:Lradio}). In Figure \ref{fig:Lradio_Rsh} we show the predicted luminosity--size relation, weighted by AGN active time, $t_{on}$. This figure illustrates that most sources have predicted sizes between 30-100 kpc, and luminosities $<10^{23} W/Hz$. In calculating the radio luminosities below, we adopt the final luminosity at age $t_{\rm on}$, rather than sampling  different points along the size-luminosity track. We adopt this simple treatment due to the limited time resolution (260 Myr) of the SAGE output. In future work, we plan to include a more accurate sampling of the luminosity--size tracks.
Furthermore, we have made the standard assumption that the synchrotron emitting electrons in the radio lobes are in the minimum energy state, i.e. there is approximate equipartition between the magnetic field and particle energy densities. Observational evidence \citep[e.g.][]{Croston2005, Shelton2011, Hardcastle2014} suggests that most sources have sub-equipartition magnetic fields, and the departure from the equipartition value depends on a number of factors, including radio source morphology. \citet{Croston2005} quote median magnetic field values that are marginally sub-equipartition, $0.7 B_{\rm eq}$. Adopting this in our work would decrease the radio luminosities by a factor $u_B^{(5+p)/4} \sim 0.5$, i.e. a factor of two lower.

\subsection{Radio AGN properties} \label{sec:RadioAGN}

Figure~\ref{fig:Lradio-Qjet} shows the relation between jet power, $Q_{\rm jet}$, and 1.4~GHz radio luminosity. Our results are consistent with that found by \citet{Cavagnolo2010} and \citet{Heckman2014} given by the solid and dashed black lines, respectively. We note that \cite{Heckman2014} estimate the jet mechanical energy of radio sources from the observed cavities and bubbles in the X-ray gas.

In Figure~\ref{fig:RLF} we present the AGN radio luminosity function (RLF) for our jet model in two redshift intervals: z=0 and z=1. This figure indicates that our jet model, using the parameters given in Table~\ref{tab:value}, produces a positive evolution of the RLF with redshift, and compares well with the observed data of \cite{Best2012} and \cite{Best2014} across the redshift range probed, including the increase in the number density of the most luminous sources. To make this figure we have assumed that all sources have the luminosity given at the end of their lifetimes, which is an underestimate (sources typically lose luminosity as they age). Our model has a number of free parameters but is calibrated only at z=0. The fact that it produces roughly the right abundances at z=1 is encouraging. Of course, the model is build from parts which approximate highly complex physical phenomena, and in the future we plan to use better ones - for example a more holistic density profile for cooling and AGN expansion. Such sophisticated approaches to predicting the radio AGN properties \citep[e.g.][]{Turner2015} are deferred to future work.

In Figure~\ref{fig:Lradio_mass} we investigate the observed radio luminosity as a function of stellar mass, again comparing our model against the data of \cite{Best2012}. Here the gray symbols show the model AGN distribution, with the line and error giving the median and scatter. 
As can be seen, our model is more aligned with the observed properties of Low Excitation Radio Galaxies (LERGs) across most of the stellar mass range plotted, which are fuelled by cooling flows, although at the lowest masses our model galaxies could be made up of a mix with High Excitation Radio Galaxies (HERGs). The \citet{Best2012} sample extends to z=0.3, which corresponds to a completeness of approximately $10^{24} W/Hz$ at 1.4 GHz in radio luminosity (using the 99 percent NVSS completeness cutoff of 3.4 mJy), and $10^{11} M_{\odot}$ in stellar mass \citep[using a SDSS r-band magnitude limit of 17.77 and the relations of][]{Bell2003}. These observations are therefore likely to be complete in stellar mass for the bulk of the bright LERG population, however not for HERGs where the observed median masses should be treated as upper limits.
	
\section{Discussion and Conclusions}
 
In this work we present a new prescription for AGN feedback that models in a more realistic way the intermittent nature of black hole accretion in the galaxy population and its resulting outflows. This update is built upon the SAGE semi-analytic galaxy formation model described in \cite{Croton2016}, which itself was an update to the model that introduced the original ``radio mode'' semi-analytic prescription in \cite{Croton2006}. Our enhancements extend to gas cooling from the hot halo to make the cooling--heating cycle self consistent, and include a new hot gas density profile that is better aligned with that observed in X-ray clusters. Overall, and after some minor parameter retuning, our additions retain the good fit SAGE has to a number of key galaxy population statistics, like the stellar mass function (for which AGN heating is critical at the high-mass end) and star formation rate density evolution, while expanding the number of observables it can make predictions for, in particular for AGN and radio galaxies. 

Unlike most semi-analytic approaches, the AGN energy injection in our model is spatially distributed.  \citet{Omma2004} pointed out that, for large radio sources, much of the AGN kinetic power can be wasted by coupling to hot gas at large cluster-centric radii, which has long cooling times. By modelling the dynamical evolution of jet-inflated structures explicitly, we address this issue.

In this new model AGN and its resulting feedback are followed through a number of key phases with time: (1) black hole accretion acts to turn on a jet, which quickly expands into the surrounding hot halo gas and inflates a cocoon; (2) after a period of time the accretion stops, and hence also the jet, which removes pressure support from the cocoon leading to its deflation (return), and ultimately dissipation; and finally (3) the galaxy then undergoes a period of quiescence, which lasts until accretion onto the black hole is re-established, after which the cycle begins again. Importantly, only during the quiet phase do we allow cooling of gas onto the galaxy, with the duration of this time set by our prescriptions for the AGN jet on, off, and cocoon return times.

Within our model gas heating from an AGN jet acts to raise the mean temperature of the hot halo by up to $\sim 20\%$ above the virial temperature (Figure~\ref{fig:Temp_hist}). When compared to the previous SAGE model, we obtain a superior match to the observed cooling luminosity--X-ray hot gas temperature relation, although we tend to under-predict the observations for the most massive clusters (Figure~ \ref{fig:Cooling_Temp}). Our more realistic cooling rates follow from the combined effect of the modified hot gas temperature and the influence that the jet inflated cavities have on the movement of this gas.

New to our model is its ability to make predictions for the properties of radio galaxies. We explore the relationship between jet power, shock radius, and radio luminosity, and compare with current observations (Figures~ \ref{fig:Lradio_Rsh} and \ref{fig:Lradio-Qjet}). The model produces a good match to both the observed radio luminosity functions from $z=0$ to $z\sim1$ (Figure~ \ref{fig:RLF}), and the local stellar mass--radio luminosity relation (Figure~\ref{fig:Lradio_mass}). For this later result, our AGN more closely align with the Low-Excitation Radio Galaxy (LERG) population, rather than the High-Excitation (HERG) population, when compared with the observations of \cite{Best2012}. Indeed,  we can match the stellar mass function in a number of ways, but to match the radio luminosity function as well we find our free parameters, as given in Table \ref{tab:value}, must be chosen to produce AGN with LERG-like properties.

However our model is not without its simplifications and areas for improvement. For example, the model only follows the active phase of jet inflated cavities and not the rising buoyant bubbles that can influence the hot gas long after the AGN has switched off (as seen from X-ray cavities with radio emission). Thus, our sources may better represent a population of FR-IIs, and our model may \emph{underestimate} the total amount of feedback done by the AGN. In addition, we ignore the possibility of sound waves distributing heating to large radii. On the other hand, we have also assumed that all cavities have simple morphology with a covering factor of 1. In reality cavities show complicated shapes, with material constantly falling back in behind. As such, these assumptions will tend to \emph{overestimate} the efficiency of feedback. \citet{Alexander2002} considered the evolution of the swept-up shell of shocked gas, and showed that the cooling time of the swept up gas depends sensitively on the thermal conductivity of the cluster. In the case that the shocked gas evolves isothermally (as would be the case for clusters in which thermal conductivity is not suppressed), radiative cooling was shown to only be significant in the centres of cool core clusters. On the other hand, the shocked gas can cool rapidly if it evolves adiabatically. Our results suggest that thermal conductivity in clusters may provide an important feedback mechanism, as previously suggested by other authors \citep[e.g.][]{Narayan2002,McCourt2013}.

Finally, it is important to note that although there is a consistency between our model predictions and observations in the $Q_{\rm jet}-L_{\rm radio}$ relation (Figure \ref{fig:Lradio-Qjet}), this relation derived from observations of X-ray cavities suffers strongly from selection effects \citep{Godfrey2016}. A relation based on dynamical models of radio AGN \citep{Turner2015} is not subject to such selection effects, but most likely provides a lower limit on the jet power for a given radio luminosity because of the assumption of an equipartition magnetic field in the radio cocoon. Inverse-Compton observations \citep[e.g.;][]{Croston2005, Shelton2011, Hardcastle2014} suggest that this assumption may overestimate the magnetic field strength by as much as a factor of four. According to Equation \ref{eqn:u_B}, a lower magnetic field would give a higher $Q_{\rm jet}$ for the same radio luminosity. 

Despite these shortcomings this paper presents a new effort to incorporate the effects of AGN jet and cavity production into a cosmological model of galaxy formation. 
Before we can make predictions for future radio  surveys  we first need to have a well constructed cosmological radio AGN model that produces a population which matches a baseline set of key statistics. The majority of our figures are calibration plots, demonstrating that our  model construction and parameter  choices by-and-large  achieve this  (for the properties considered at least). 

This model is able to produce catalogues of both normal and radio galaxies across a wide range of mass scales and environments, describing their evolution from high redshift down to the present day. Such catalogues offer a valuable resource for survey teams as they analyse current data and plan for the future. 
Extensions to our current work will focus on the geometry of bubble evolution (e.g. how do bubbles ``pancake'' and sweep out the gas), how gas refills behind the bubble as it evolves, and how the local magnetic field helps stabilise the bubble and delays gas from returning. To achieve such detail we will employ more sophisticated high resolution numerical simulations, with the results then generalised and folded into our semi-analytic model, expanding its predictive power and usefulness to the community.

\section*{Acknowledgements}

We thank the anonymous referee for their constructive comments and suggestions which helped to improve the paper. MR is grateful to Swinburne University and its Centre for Astrophysics and Supercomputing for hosting his stay during which much of this work was carried out. SS thanks the Australian Research Council for an Early Career Fellowship (DE130101399).

The Semi-Analytic Galaxy Evolution (SAGE) model that served as a basis for this work is a publicly available codebase that runs on the dark matter halo trees of a cosmological N-body simulation. It is available for download at https://github.com/darrencroton/sage.
The Millennium Simulation on which the semi-analytic model was run was carried out by the Virgo Supercomputing Consortium at the Computing Centre of the Max Plank Society in Garching. It is publicly available online at http://www.mpa-garching.mpg.de/Millennium/ through the German Astrophysical Virtual Observatory.

	\bsp
	\label{lastpage}

\end{document}